# Reemergence of superconductivity in pressurized quasi-one-dimensional superconductor $K_2Mo_3As_3$

Cheng Huang[1,2]*, Jing Guo[1,4]*, Kang Zhao[1,2]*, Fan Cui[1,2], Shengshan Qin[1,2], Qingge Mu[1], Yazhou Zhou[1], Shu Cai[1,2], Chongli Yang[1], Sijin Long[1,2], Ke Yang[3], Aiguo Li[3], Qi Wu[1], Zhian Ren[1,2], Jiangping Hu[1,2] and Liling Sun[1,2,4]†

[1]Institute of Physics and Beijing National Laboratory for Condensed Matter Physics, Chinese Academy of Sciences, Beijing 100190, China

[2]University of Chinese Academy of Sciences, Beijing 100190, China

[3]Shanghai Synchrotron Radiation Facilities, Shanghai Institute of Applied Physics, Chinese Academy of Sciences, Shanghai 201204, China

[4]Songshan Lake Materials Laboratory, Dongguan, Guangdong 523808, China

Here we report a pressure-induced reemergence of superconductivity in recently discovered superconductor $K_2Mo_3As_3$, which is the first experimental case observed in quasi-one-dimensional superconductors. We find that, after full suppression of the ambient-pressure superconducting (SC-I) state at 8.7 GPa, an intermediary non-superconducting state sets in and prevails to the pressure up to 18.2 GPa, however, above this pressure a new superconducting (SC-II) state appears unexpectedly. High pressure x-ray diffraction measurements demonstrate that the pressure-induced dramatic change of the lattice parameter $c$ contributes mainly to the emergence of the SC-II state. Combined with the theioretical calculations on band strcture, our results suggest that the reemergemce of superconductivity is associated with the change of the complicated interplay among different orbital electrons, driven by the pressure-induced unisotropic change of the lattice.

The discoveries of the copper oxide and iron-based high-$T_c$ unconventional superconductors have generated considerable interest over the past 30 years due to their unusual superconducting mechanism and great potential for application. A common feature of these superconductors is that they contain $d$ orbital electrons in their quasi-two-dimentional lattice that results in intriguing superconductivity and other exotic properties [1-6]. Recently, a family of CrAs-based superconductors with superconducting transition temperatures ($T_c$) of 2.2 - 8.6 K were found in $X_2Cr_3As_3$ (X=Na, K, Rb, Cs) compounds, which possess the features of containing $d$ orbital electrons in a quasi-one-dimensional lattice [7-10] and many unconventional properties [11]. Soon after, a new family of quasi-one-dimmensional MoAs-based superconductors $Y_2Mo_3As_3$ (Y= K, Rb, Cs) with $T_c$s from 10.4 K to 11.5 K was discovered [12-14]. The properties of the upper critical field and specific-heat coefficient of the $Y_2Mo_3As_3$ superconductors are similar to those of $X_2Cr_3As_3$.

The crystal structure of the $Y_2Mo_3As_3$ superconductors is same as that of $X_2Cr_3As_3$, crystallized in a hegxagonal unit cell without an inversion symmetry [12]. The alkali metallic atoms serve as the charge reservoir and the Mo-As chains are responsible for their superconductivity [12]. Since these superconductors host non-centrosymmetric structure that usually connects to the unconventional pairings and exotic physics, they have received considerable attentions in the field of superconductivity researches [15-19].

Pressure tuning is a clean way to provide significant information on evolution among superconductivity, electronic state, and crystal structure without changing the chemical composition, and eventually benefits for a deeper understanding of the underlying physics of the puzzling state emerged from ambient-pressure materials

[2,20-24]. Earlier high pressure studies on $X_2Cr_3As_3$ (X=K and Rb) below ~4 GPa found that its superconductivity can be dramatically suppressed by external pressure [25,26], however, high pressure studies on the $K_2Mo_3As_3$ superconductor is still lacking. In this study, we performed the *in-situ* high pressure transport measurements on the $K_2Mo_3As_3$ sample up to 50 GPa to know what is new for this compound under high pressure.

The polycrystalline (wire-like) samples, as shown in the inset of Fig. 1a, were synthesized using a conventional solid-state reaction method, as described in Ref. [12]. High pressure was generated by a diamond anvil cell made from a Be-Cu alloy with two opposing anvils. Diamond anvils with 300 culets (flat area of the diamond anvil) were used for the experiments. In the resistance measurements, we used platinum foil as electrodes, rhenium plate as gasket and cubic boron nitride as insulating material. The four-probe method was employed to determined pressure dependence of superconducting transition temperature. In the high-pressure *ac* susceptibility measurements, the sample was surrounded by a secondary coil (pickup coil), above which a field-generating primary coil was wounded [21, 27]. High-pressure angle dispersive x-ray diffraction (XRD) measurements were carried out at beamline 15U at the Shanghai Synchrotron Radiation Facility. A monochromatic x-ray beam with a wavelength of 0.6199 Å was adopted. The pressure was determined by ruby fluorescence method [28]. Given that the sample reacts with the pressure transmitting mediums under pressure, no pressure medium was adopted in all high-pressure measurements.

Figure1a shows the temperature dependence of the electrical resistance of the ambient-pressure sample with an onset $T_c$ of about 10.4 K, in good agreement with the previous report [12]. Figure 1b-1d show the high pressure results. It is seen that the resistance at 1.2 GPa shows a remarkable drop starting at ~9.2 K and reaches zero at ~ 4.5 K (inset of Fig. 1b), indicating that the applied pressure decreases the onset $T_c$ from 10.4 K to 9.2 K. Upon further increasing pressure to 1.5 GPa, the sample loses its zero resistance and its $T_c$ shifts to lower temperature, reflecting that the superconductivity of this polycrystalline sample is highly sensitive to the applied pressure. Then, $T_c$ decreases continuously with a rate of $dT_c/dP$= -1.19 K/GPa until cannot be detected at 8.7 GPa down to 1.6 K (Fig. 1b). The non-superconducting state persists to the pressure of 18.2 GPa, at which unexpectedly another remarkable resistance drop is found at ~ 4.6 K (Fig. 1c). This drop becomes more pronounced with further compression (Fig. 1d), and plunges about 92% at 47.4 GPa. Moreover, we find that the onset temperature of the new resistance drop shifts to high temperature with increasing pressure initially (the inset of Fig. 1d), reaches a maximum (~8.1 K) at ~ 30 GPa and saturats up to ~38 GPa. By applying higher pressure to 47.4 GPa, the temperature of the drop displays a slow decline (inset of Fig. 1d).

To confirm whether the new resistance drop observed in $K_2Mo_3As_3$ is related to a superconducting transition, we applied magnetic field to the sample subjected to 19.6 GPa and 41.8 GPa, respectively (Fig. 2a and 2b). It can be seen that this new resistance drop shifts to lower temperature with increasing magnetic field and almost suppressed under the magnetic field of 3.5 T and 4.0 T for the compressed sample at 19.6 GPa and 41.8 GPa, respectively. These results indicate that the new resistance drop should be resulted from a superconducting transition. The alternating-current (*ac*) susceptibility measurements were also performed for the sample subjected to 20 GPa - 44 GPa down to 1.5 K, the lowest temperature of our instrument, but the diamagnetism is not detected. By our analysis, the failure of the measurements on the diamagnetism may be related to the low volume fraction of the pressure-induced superconducting phase.

We extract the field ($H$) dependence of $T_c$ for $K_2Mo_3As_3$ at 19.6 GPa and 41.8 GPa (Fig. 2a and 2b) and plot the $H(T_c)$ in Fig. 2c. The experimental data is fitted by using Ginzburg-Landau (GL) formula, which allows us to estimate the values of the upper critical magnetic field ($H_{C2}$) at zero temperature: 4.0 T at 19.6 GPa and 4.8 T at 41.8 GPa (Fig. 2c). Note that the upper critical fields obtained at 19.6 GPa and 41.8GPa are lower than their corresponding Pauli paramagnetic limits (10.3 T and 14.8T, respectively), suggesting that the nature of the pressure-induced superconducting state may differ from that of the initial superconducting state.

To investigate whether the observed reemergence of superconductivity in pressurized $K_2Mo_3As_3$ is associated with the pressure-induced crystal structure phase transition, we performed *in-situ* high pressure XRD measurements. The XRD patterns collected at different pressures are shown in Fig. 3. No structure phase transition is observed under pressure up to 51.6 GPa. And all peaks shift to higher angle due to the shrinkage of the lattice, except for the (002) peak. It is found that the (002) peak, which is realted to the parameter $c$, shifts toward to the lower angle starting at 8.8 GPa. Upon further compression, it becomes more pronounced. We propose that the left shift of the (002) peak may be a consequence of the pressure-induced elongation of the polycrystalline wire-like samples (the direction of the wire length is the $c$-axis of the $K_2Mo_3As_3$ crystal lattice) due to their preferred orientation under pressure (see the right panel of Fig. 3). At higher pressure, the wire-like samples aline themselves perpendicular to the pressure direction applied.

We summarise our results in Fig. 4a. The pressure-$T_c$ phase diagram clearly reveals three distinct superconducting regions: the initial superconducting state (SC-I), intermediary non-superconducting state (NSC) and the pressure-induced superconducting state (SC-II). In the SC-I region between 1 bar and 8.7 GPa, $T_c$ is suppressed with applied pressure, and not detectable above 8.7 GPa. In the SC-II region, $T_c$ increases with pressure and reaches the maximum (8.1 K) at 30 GPa. Upon further compression, $T_c$ shows slow a slight decline. This is the first observation of reemerging superconductivity in one-dimensional superconductors, to the best of our knowledge.

We extract the lattice parameters and volume as a function of pressure, and summarise these results in Fig. 4b. It is found that the lattice constant *a* displays a decrease monotonously with pressure, while the lattice constant *c* shows a complicated relation with pressure applied. In the SC-I region, the parameter *c* shrinks upon increasing pressure, but it unusually expands in the NSC region. At the pressure of 18.2 GPa, the superconductivity reemerges and the lattice constant *a* and *c* decrease simultaneously with pressure again as what is seen in the SC-I region, implying that the pressure-induced elongation effect on the wire-like sample is satuated. The pressure dependence of the volume for $K_2Mo_3As_3$ is shown in the inset of the Fig. 4b, displaying that the sample volume remains almost unchanged due to the increase of parameter *c* and the decrease of parameter *a* concurrently with elevating pressure in the NSC region. The strong correlation between the lattice parameters and the superconductivity in pressurized $K_2Mo_3As_3$ suggests that the remarkable change of parameter *c* may paly an important role for the development of the SC-II state.

To understand the underlying correlation between $T_c$ and the electronic state in $K_2Mo_3As_3$ further, we performed the first-principles calculations on its electronic structure, based on our XRD results, by using the projector-augmented wave (PAW) method (see the Supplemental Material [29]). We find that the percentage of the density of states (P-DOS) at Fermi level for $K_2Mo_3As_3$ is dominated by the electrons from the $d_{xy}$ and $d_{x^2-y^2}$ orbitals in the pressure range investigated, with a secondary contribution from the $d_{z^2}$, $p_x$ and $p_y$ orbitals, and the P-DOSs of the $p_z$, $d_{xz}$, $d_{yz}$ and *s* orbitals are relatively small (see the Supplemental Material [29]). We note that, in the NSC and SC-II regions, the P-DOSs of the $d_{xy}$, and $d_{x^2-y^2}$ orbitals decrease continuously with elevating pressure over the experimental range investigated, however, the change trend of the P-DOSs contributed by the $d_{z^2}$ and the $p_x$ as well as the $p_y$ orbitals displays differently. In the NSC region, the P-DOS of the $d_{z^2}$ orbital exibits a remarkable increase, while that of the $p$ orbitals shows a slow decline (Fig. 4c). As the P-DOSs of the $d_{z^2}$ orbital and $p$ orbitals reach a maximum and minimum respectively, SC-II state appears. Note that the $T_c$ value of the SC-II state increases upon compression in the pressure range of 18.2 GPa – 27 GPa, just where the P-DOS of the $d_{z^2}$ orbital displays a decrease again. Meanwhile, the P-DOSs of the $p_x$ and $p_y$ orbitals appear an increase in the same pressure range. Further compression from 27 GPa to 40 GPa, $T_c$ of the SC-II state stays almost constant, and the corresponding P-DOSs of the $d_{z^2}$ and $p$ orbitals show a small change (Fig. 4c). These results suggest that the emergence of the SC-II state in $K_2Mo_3As_3$ and its $T_c$ change with pressure are the consequence of the interplay among the different orbital electrons.

In conclusion, the pressure-induced reemergence of the superconductivity is observed for the first time in the qausi-one-demensional superconductor $K_2Mo_3As_3$. An intimate correlation between $T_c$'s of the ambient-pressure and high-pressure superconducting states, lattice parameters and the density of state contributed by $d_{z^2}$, $p_x$ and $p_y$ orbitals have been revealed. We find that the initial superconducting state (SC-I) is suppressed by pressure at 8.7 GPa, and then an intermediary non-superconducting (NSC) state sets in and stabilizes up to 18 GPa. Subsequently, a new superconduting state (SC-II) emerges and prevails up to 47.4 GPa. Our synchrotron x-ray diffraction results indicats that the reemergence of superconductivity is not associated with any crystal structure phase transition. In combination of theoretical calculations on the band strcture, our results suggest that the appearance of the SC-II state found in this material is a consenquence of the dramatic interplay among different orbital electrons due to the pressure-induced lattice change. We hope that the results found in this study will shed new light on understanding the correlation among superconductivity, electronic and lattice structures in unconventional quasi-one dimentional superconductors.

**Acknowledgements**

We thank Prof. V. A. Sidorov for useful discussions. The work was supported by the National Key Research and Development Program of China (Grant No. 2017YFA0302900, 2016YFA0300300 and 2017YFA0303103), the NSF of China (Grants No. U2032214, 12004419 and 12074414) and the Strategic Priority Research Program (B) of the Chinese Academy of Sciences (Grant No. XDB25000000). J. G. is grateful for support from the Youth Innovation Promotion Association of the CAS (2019008).

*contributed equally to this work.
†To whom correspondence may be addressed. Email: llsun@iphy.ac.cn

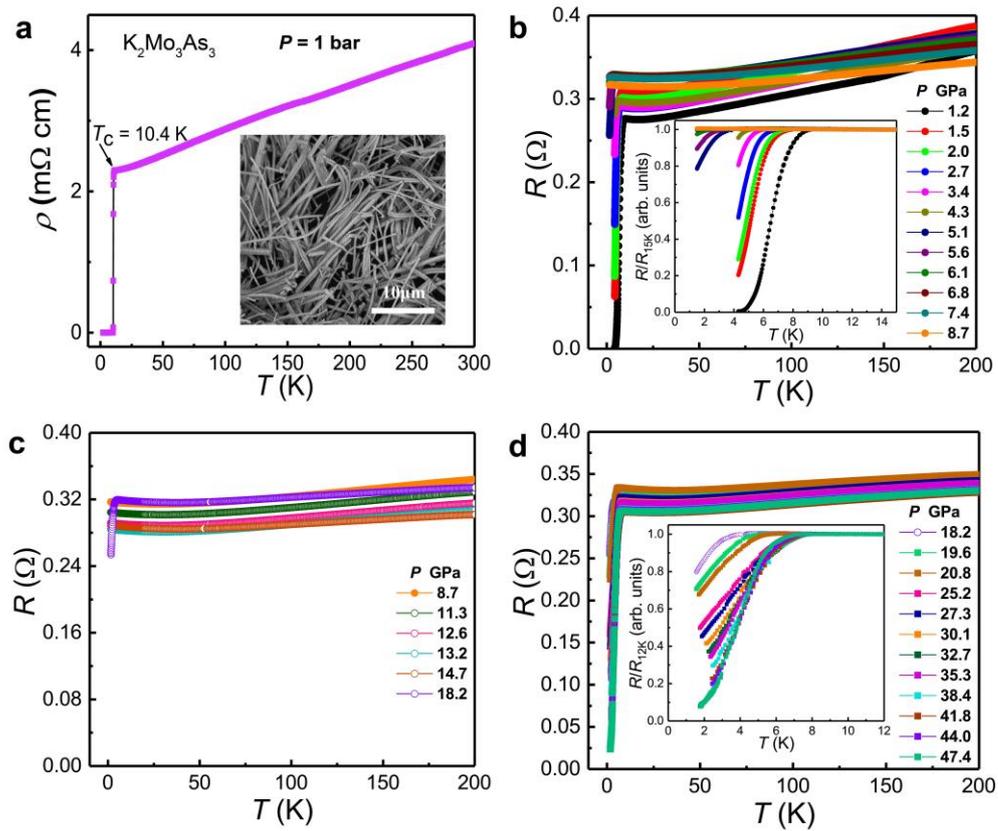

Fig. 1 (a) Resistance as a function of temperature for quausi-one-dimensional superconductor $K_2Mo_3As_3$ at ambient pressure and its polycrystallined-sample's image taken by a scanning electron microscope (inset). (b)-(d) Temperature dependence of the resistance in the pressure range of 1.2 GPa -8.7 GPa, 8.7-18.2 GPa, and 18.2-47.4 GPa, respectively. The insets of figure (b) and (d) display enlarged views of the the resistance in the low temperature regime.

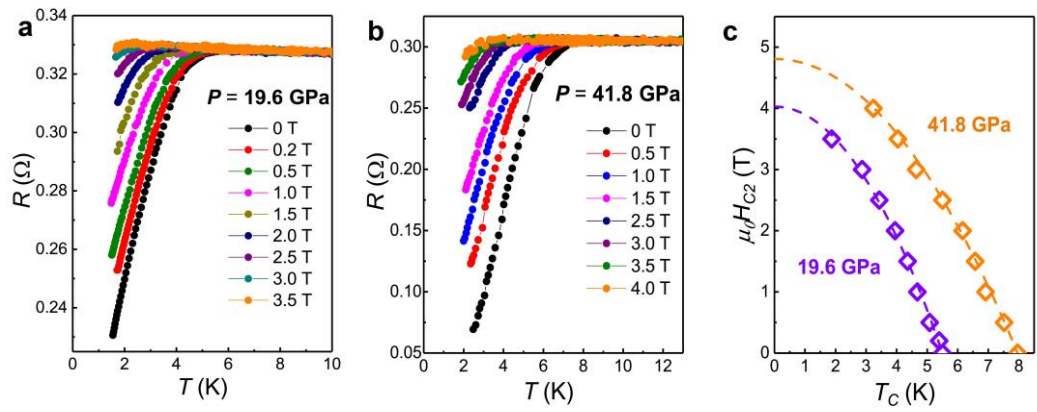

Fig. 2 Temperature dependence of resistance under different magnetic fields for K$_2$Mo$_3$As$_3$ measured at 19.6 GPa (a) and 41.8 GPa (b), respectively. (c) Plot of superconducting transition temperature $T_c$ versus critical field ($H_{C2}$) for K$_2$Mo$_3$As$_3$ at 19.6 GPa and 41.8 GPa, respectively. The dished lines represent the Ginzburg-Landau (GL) fits to the data of $H_{C2}$.

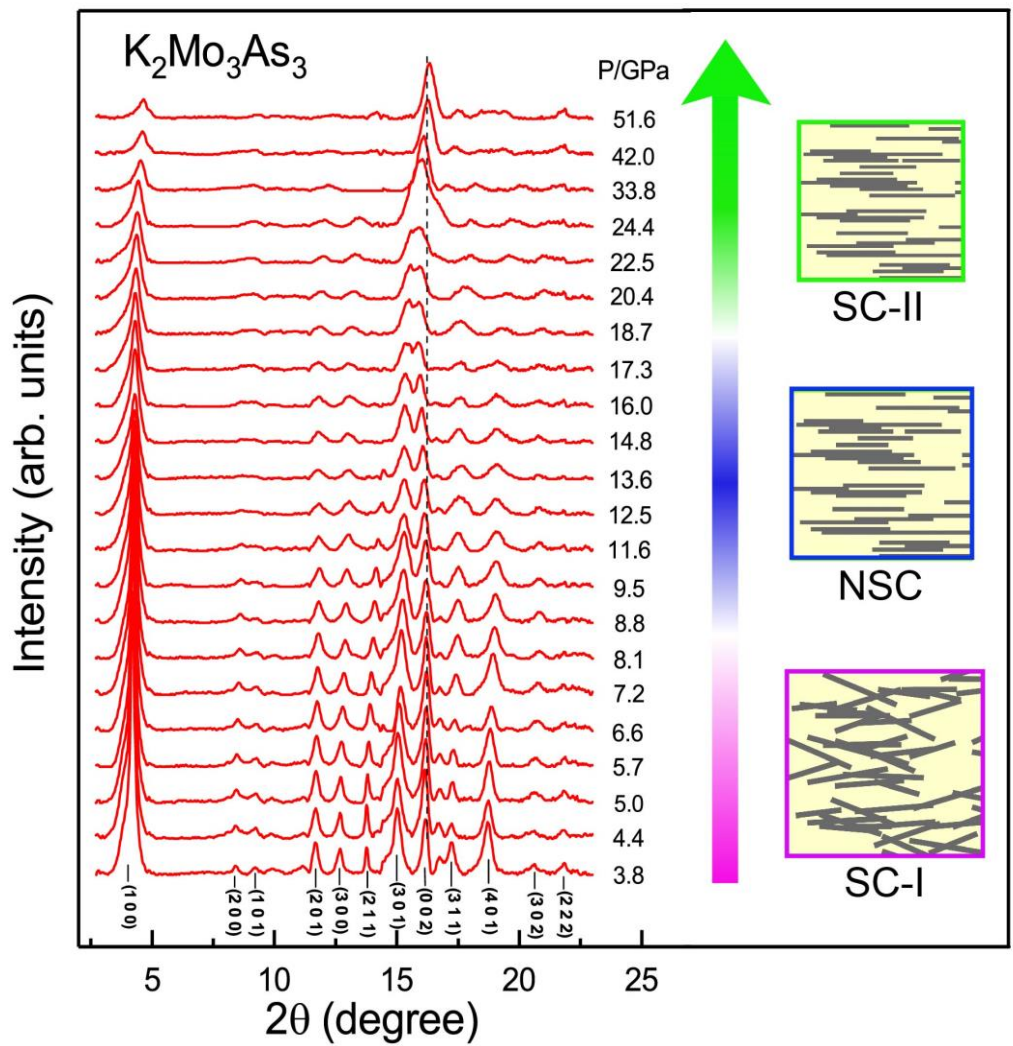

Fig. 3 X-ray diffraction patterns collected in the pressure range of 3.8 – 51.6 GPa for K$_2$Mo$_3$As$_3$. The right panel schematically shows the evolution of the preferred orientation of the wire-like samples under pressures. SC-I and SC-II stand for the ambient-pressure superconducting state and the reemergence supercnducting states, respectively. NSC represents non-superconducting state.

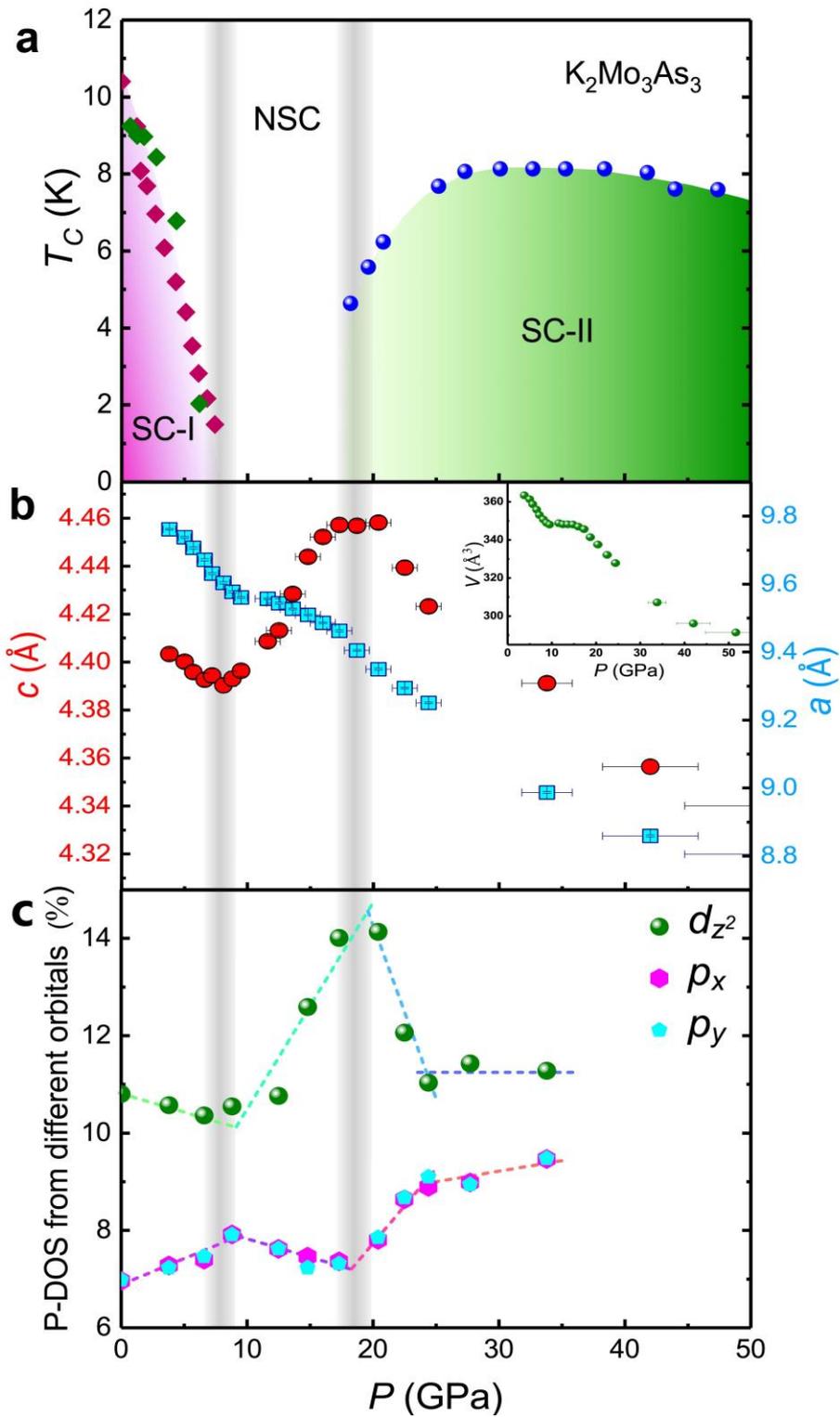

Fig 4 Superconductivity, crystal and electronic structure information for the qausi-one-dementional superconductor K$_2$Mo$_3$As$_3$. (a) Pressure-Temperature phase diagram. (b) Pressure dependence of lattice parameters and volume (see inset). (c) Plots of percentage of the density of state (P-DOS) contrinuted by the $d_z^2$, $p_x$ and $p_y$ orbitals versus pressure.